\documentstyle{mn}

\catcode`\"=\active\let"=\"

\def\pau{P. Amaro-Seoane}
\def\msol{{\rm M}_\odot}
\def\lsol{{\rm L}_\odot}

\title[The loss-cone problem in dense nuclei]
{The loss-cone problem in dense nuclei}
\author[\pau~and R. Spurzem]
{\pau ~and R.~Spurzem \\
Astronomisches Rechen-Institut, M"onchhofstra\ss e 12-14,
   D-69120 Heidelberg, Germany \\
e-mail: pau, spurzem@ari.uni-heidelberg.de}

\begin{document}

\maketitle

\begin{abstract}
We address the classical problem of star accretion onto a
supermassive central gaseous object in a galactic nucleus.
The resulting supermassive central gas-star object is assumed to
be located at the centre of a dense
stellar system for which we use a simplified model consisting of a
Plummer model with an embedded density cusp using stellar point 
masses. From the number of stars belonging to the loss-cone, which
 plunge onto the central object on elongated orbits from outside,
 we estimate the accretion rate taking into account a possible anisotropy of the
 surrounding stellar distribution. 
The total heating rate in the
 supermassive star due to the loss-cone stars plunging onto it
 is estimated. This semi-analytical study, revisiting and
 expanding classical paper's work, is a starting point of
future work on a more detailed study
 of early evolutionary phases of galactic nuclei.
 It merits closer examination, because it
 is one of the key features for the link
 between cosmology and galaxy formation.
\end{abstract}


\section{General Introduction}

 Supermassive black holes (SMBH hereafter) lurking in centres of dense galactic nuclei,
 accreting stars and gas, provide under certain conditions the most
 powerful sources of energy in our visible universe, the quasars. 
 That rather exotic idea in early time (Salpeter 1964, Zel'dovich 1964,
 Lynden-Bell 1969, Lynden-Bell \& Rees 1971, Rees 1984, Begelman,
 Blandford \& Rees 1984)
 has become common sense nowadays. Not only our
 own galaxy harbours a few million-solar mass black hole 
(Genzel et al. 2000, Genzel 2001, Ghez et al. 2000)
 but also many of other non-active galaxies show kinematic
 and gasdynamic evidence of these objects (Magorrian et al. 1998). 
 The question
 how the black hole continues to grow, how it influences the stellar
 and gas distribution around itself, was intensively studied more than
 two decades ago (Frank \& Rees 1976, Bahcall \& Wolf 1976, Marchant \& Shapiro
 1980). 
 Since it became clear relatively
 early that most supermassive black holes cannot be formed fast enough from
 stellar mass seed black holes in nuclei (Duncan \& Shapiro 1982, 1983, but
 see also Lee (1995) for a somewhat differing view), they must be
 formed during the galaxy formation process directly, which is linked
 to cosmological boundary conditions. Rees (1984, 1996) argued that 
 galactic nuclei in their formation process inevitably produce a dense
 core consisting of a star-gas system or a cluster of compact stellar
 evolution remnants, both ultimately collapsing to a supermassive black hole.
 Supermassive stars (SMS from now onwards) as such have been studied (Hoyle et al. 1964,
 Fricke 1973, 1974, Fuller et al. 1986) or dense supermassive star-gas 
 composite objects as a transient progenitor
 of a SMBH (Hara 1978, Hagio 1986, Fuller et al. 1986,
 Langbein et al. 1990). Finally, the stability of compact dense star clusters
 was examined (Zel'dovich \& Podurets 1965, Quinlan \& Shapiro 1990). All
 these papers as a common feature conclude that, provided the central
 object, star cluster, SMS or a mixture of both, becomes smaller than
 a certain critical radius, it is able to undergo catastrophic collapse
 in a dynamical time scale due to an instability caused by Post-Newtonian
 relativistic corrections of hydrostatic equilibrium. The question, however,
 whether and how that final unstable state can be reached, is much less
 clear. Angular momentum of the protogalaxy or its dark matter halo,
 self-enrichment during the dissipative collapse providing opacity through
 lines which prevents collapse as compared to radiation driven expansion,
 and star-gas interactions heating the central massive gas object could all
 at least for some time prevent the ultimate collapse. Given the complex
 physical nature of the interstellar matter, star formation, and stellar
 interactions alone this is a complicated question and the conditions
 under which a supermassive object can form in a spherical, isolated
 star-forming and collapsing gas cloud, rotating or not, has to our 
 knowledge never been exhaustively studied and answered (see however,
 some pioneering approaches such as Colgate 1967, Sanders 1970, Spitzer \& Stone 1967,
 Spitzer \& Saslaw 1966, Langbein et al. 1990, Quinlan \& Shapiro 1990). 
 The question has gained even more complexity, since we now know
 that the baryonic matter of galaxies collapses in their dominating dark
 halo, and that most galaxies and their dark halos experience merging
 with other dark haloes and large and small galaxies during the hierarchical
 gravitational structure formation (cf. e.g. Kauffmann et al. 1999a, b,
 Diaferio et al. 1999).
 This has led to another type of study
 of black hole statistics: due to e.g. Kormendy \& Richstone (1995) and
 Magorrian et al. (1998) the black hole masses are well correlated with
 the bulge masses of their mother galaxies, and Gebhardt et al. (2000) find
 a correlation with the host galaxy velocity dispersion. These correlations
 support the idea that black hole formation is linked to galaxy formation.
 Such idea has been studied earlier, using a Press-Schechter hierarchical
 structure formation model (Efstathiou \& Rees 1988), and the results
 can be checked
 against the quasar luminosity function (Small \& Blandford 1992,
 Eisenstein \& Loeb 1995, Haehnelt \& Rees 1993,
 Haehnelt et al. 1998, Boyle et al. 2000). On the basis of a statistical
 analysis using semi-analytical merger trees of galaxy building in a
 hierarchical structure formation picture Haehnelt \& Kauffmann (2000) are
 able to reproduce the observed correlations using simple scaling
 relations for many of the in detail unknown physical processes,
 such as star formation, baryonic matter collapse in the halos, to mention
 only two examples. The detailed physics and parameters, how these processes
 work in a self-consistent model of black hole formation, however,
 are much less understood. So we lack any idea, what are the signatures
 of the black hole formation process in the morphology and kinematics of
 the innermost core and cusp regions, and to what extent they survive the
 merging history. Brave attempts to advance modelling in that domain
 (Rauch 1999) demonstrate in our view more the problems which still prevail
 originating from the large dynamical range of the problem and the 
 complexity of the treatment of relaxation in a stellar system, rather
 than that they provide much reliable new insight.

 Galaxy merging poses another serious problem due to the possibility that
 it can lead to two or more black holes in one nucleus, and the structure
 and kinematics is critically dependent on the evolution and possible
 gravitational radiation merger of the resulting black hole binary. 
 For that problem direct $N$-body modelling is the only available 
 possibility here using special GRAPE supercomputers (Makino \& Ebisuzaki 1996,
 Makino 1997), general purpose supercomputers (Merritt, Cruz \& Milosavljevic
 2000), or a suitable hybrid method between direct and approximate
 $N$-body codes (Hemsendorf, Sigurdsson \& Spurzem 2001). From these models
 it is yet unclear how fast in the real system dynamical friction,
 stochastic three-body interactions and external perturbations work together
 to produce eventually a single black hole again.

 On the other hand, due to the ever increasing
 observational capabilities with ground and space based telescopes we get
 more and more detailed dynamical and photometric data of the structure of
 stellar systems around black holes. Therefore, we find it worthwhile to
 reconsider with present day numerical possibilities and increased 
 knowledge about galaxy formation and evolution a detailed study of
 the evolution of dense star clusters, with gas, forming an SMS and
 its further evolution

 In this paper we first reconsider the problem of a SMS in a dense star
 cluster and study with semi-analytic means its growth by star-gas
 interaction and trapping of stars, using a fixed, approximate density 
 and velocity dispersion profile of the surrounding star cluster. The
 size of the loss-cone and energy generation rate due to star-gas
 interactions are derived in a generalization of the ``classical''
 determinations by Frank \& Rees (1976), da Costa (1981) and Hara (1978).
 Subsequent work will incorporate these results in time-dependent numerical
 chemodynamical models of nuclei, first in spherical symmetry and later
 also for other non-symmetric configurations.

\section{Introduction: The loss-cone, a review and further details }

Since the SMS scheme embraces the
 black hole accretion problem, we will have first a look at this
 problem in order to extract from it the global concept for the SMS.
Thus, in this section we discuss the consequences from the 
astrophysical and dynamical point of view of the presence of
 a black hole (or a massive compact central object, from now onwards just
 BH) in a dense stellar object
 in about a relaxation time. We consider the steady-state
 distribution and consumption of stars orbiting a massive
 object at the centre of a spherical, stellar system.

The distribution of stars is determined by the relaxation
 processes associated with gravitational stellar encounters
 and by the consumption of low angular momentum stars which
 pass within a small distance of the central mass.
Stars whose orbits carry them within the tidal radius $r_{t}$
 of the BH will be tidally disrupted. The {\sl peribarathron}
 (distance of closest approach to the BH) is determined by
 the specific orbital angular momentum {\sl L} and by the BH mass.
There are situations in which the stars that have a radially
 elongated orbit and a low angular momentum pass close by the
 system centre and interact with the massive central object.
 In such a situation it is interesting to evaluate the density
 of those stars whose angular momenta are limited by a superior
 $L_{\rm min}$; that is, the stars which belong to a defined region
 in the velocity-space.

Stars at a position $r$ whose velocities are limited by a superior
 limit $v_{lc}(r)$ and, consequently, with an angular momentum 
$L < L_{\rm min}=rv_{lc}$, have orbits that will cross the tidal
 radius of the central BH in their motion. They will be disrupted
 due to the tidal forces and then they are lost for the stellar
 system. Such stars are said to belong to the {\em loss-cone},
 since they are lost for the stellar system. The loss-cone is
 depleted in a crossing time $t_{\rm cross}=r/ \sigma$, where 
$\sigma$ is the 1D velocity dispersion.

The diffusion of stars into this loss-cone has been studied by
 Frank and Rees (1976) and by Lightman and Shapiro (1977). We
 will talk about a ``critical radius'' within which stars on 
orbits with $r\leq r_{\rm crit}$ diffuse. Inside they are 
swallowed by the BH, after being scattered to low angular momentum
 {\sl loss-cone} orbits. 

If there is a central point mass ${\cal M}_{\bullet}$, such that 
${\cal M}_{\bullet}$ $\gg$ $m_{*}$,
 then its potential well will affect the stellar velocity field out to a distance
\begin{equation}
r_{h}= G{\cal M}_{\bullet}/\sigma^2.
\end{equation}
\noindent
This expression gives us the influence radius of the central
 object. $G$ is the gravitational constant.
The star gets disrupted whenever the work exerted over the star
 by the tidal force exceeds its own binding energy. 
If we compute the work exerted over the star by the BH we can get
 an expression for the tidal radius,
\begin{equation}
r_{\rm t}=\Bigg[\frac{2}{3} (5-n) \frac{{\cal M}_{\bullet}}{m_{*}}\Bigg]^{1/3} r_{*},
\end{equation}
\noindent
where $n$ is the polytropic index of the star, $m_{*}$ the mass of the
 star and $r_{*}$ its radius.

For solar-type stars it is (considering a $n=3$ polytrope)

\begin{equation}
r_{\rm t} \simeq 4.5\cdot 10^{-8} \bigg( \frac {{\cal M}_{\bullet}}{\msol} \bigg)^{1/3}~{\rm pc}.
\end{equation}

\subsection{Loss-cone phenomena}

Frank and Rees (1976) studied how a stationary stellar density profile
 around a massive star accreting BH looks like. They found that the density
 profile follows a power-law within the region where the gravity of the massive
 star dominates the self-gravity of the stars,
\begin{center}
\begin{equation}
\rho \propto r^{-7/4}.
\end{equation}
\end{center}
This was followed by intensive numerical studies by other authors
 (cf. e.g. Shapiro \& Marchant 1978, Marchant \& Shapiro 1979, 1980
 and Shapiro 1985) which are all in agreement with the first work
 of Frank and Rees. 

The replenishment of the loss-cone happens thanks to the small-angle
 gravitational encounters in a timescale which is for the most real
 stellar systems slower than the dynamical processes.

We have loss-cone effects also in the neighbourhood of a massive
 central gas-formed object, an SMS: Stars with such orbits enter the gas-formed
 central object and lose kinetic energy if their density is high enough.
 Nevertheless, such stars will not disappear from the stellar system just
 by one crossing of the central object, as it happens for the BH problem. In
 this scenario the stars lose their energy in each crossing and their orbits
 come closer and closer to the central object until they are ``trapped'' in it,
 their orbits do not extend further than the massive central object radius 
(confined stars). This process was described qualitatively by Hara (1978),
 da Costa (1981) and Hagio (1986). Da Costa suggested the name ``dissipation-cone''.
 However, we will keep the loss-cone term for it is the most commonly found in the related
 literature. We have to take into account that the meaning is that of a defined 
region of the velocity space at the position $r$, even though there is no quick
 loss.

Frank and Rees (1976) derived the following expression for the diffusion
 angle (the mean deviation of a star orbit in a dynamical time $t_{\rm dyn}$):
\begin{equation}
\theta_{D} \simeq \sqrt{\frac {t_{\rm dyn}}{t_{\rm relax}}}.
\end{equation}
\noindent
where $t_{\rm dyn}$ and $t_{\rm relax}$ are the dynamical and relaxation times; namely,
\begin{equation}
t_{\rm dyn}=r/ \sigma_{r}(r)
\end{equation}
\begin{equation}
t_{\rm relax}= \frac {9}{16 \sqrt {\pi}} \frac {\sigma ^3 }{G^2m \rho(r) \ln(\eta N)}.
\end{equation}
\noindent
In this last expression (Larson 1970), $\rho$ is the mean stellar mass density,
 $N= \frac{4}{3}\pi n_{c}r_{c}^3$ is the total particle number and $\eta$ is a
 parameter of order unity which is set to be 0.4 (Spitzer 1958) or 0.11
 (Giersz and Heggie 1994); its exact value cannot be defined easily and
 depends on the initial model and the anisotropy. Here we use $\eta=0.11$.

Now we look for a condition at a place $r>r_{h}$ for a star to touch or
 to cross the influence radius of the central object within a crossing
 time. For this aim we look now for the amount of stars which reaches
 the central influence radius with an unperturbed orbit. Unperturbed
 here means that the star orbit results from the influence of the
 gravitational potential and from that of the rest of the stars and of the central
 object, and it is not affected by the local, two-body, small- or
 big-angle gravitational encounters. We envisage then the average
 part of the gravitational potential, whereas the random component
 due to the individual behaviour of the stars will be neglected.

To define the loss-cone angle we say that a star belongs to this 
cone when its  distance to the peribarathron (which depends on the
 orbit we have, i.e., on the energy $E$ and angular momentum) is
 less or equal to the tidal radius, $r_{p}(E,L) \leq r_{t},~ \theta
 \leq \theta_{lc}$.

There is a maximum $\theta$ for which the peribarathron radius is
 less than or equal to the tidal radius. We define this as $\theta_{lc}$ 
(where the subscript ``lc'' stands for loss-cone).

We derive the $v_{lc}(r)$ using angular momentum and energy
 conservation arguments. We just have to evaluate it at a general
 radius $r$ and at the tidal radius $r_{t}$, where the tangential
 velocity is maximal and the radial velocity cancels:
\begin{equation}
v_{lc}(r)= \frac {r_{t}}{\sqrt {r^2-r_{t}^2}} \cdot
 \sqrt {2[ \phi (r_{t}) - \phi (r)] + v_{r}(r)^2}.
\end{equation}
\subsection{The critical radius}

It is interesting to evaluate a certain radius which Frank and Rees (1976)
 introduced by defining the ratio $\xi:= \theta_{lc}/ \theta_{D}$. When
 $\xi =1$, then $\theta_{lc}=\theta_{D}$, and this corresponds to
 a ``critical radius'', $r_{\rm crit}$, if there is only one radius with 
this condition. Inside the critical radius (i.e. $\xi >1$, $\theta_{lc}>\theta_{D}$)
 stars are removed on a $t_{\rm dyn}$. For larger radii (i.e. $\xi <1$, 
$\theta_{lc}<\theta_{D}$) we cannot talk about a ``loss-cone''
 because this $\theta_{D}$ corresponds to the variation of $\theta$ within
 a  $t_{\rm dyn}$, and this is the required time for the star within the
 loss-cone to plunge onto the BH if the orbit is unperturbed. The angle variation
 happens sooner than the required time for stars to sink into the loss-cone.
 If $\theta_{D}>\theta_{lc}$, loss-cone stars can get in and out of the
 loss-cone faster than they could reach the central object.

\section{The loss-cone contribution to the heating rate of the gas}

\subsection{Introduction}
Dissipation of the stellar kinetic energy of a star plunging onto the SMS and
 suffering from the drag force leads to a heating of the SMS. Another
 possible consequence of the local star-gas interaction is the formation
 of massive stars within the cloud due to the accretion of ambient gas
 (Da Costa 1979). This could increase the supernova rate and be an important
 source of energy. Here we will assume an equilibrium for the SMS for the
 timescales of interest.

The star distribution will be affected at large radii ($r \gg {\cal R}_{s}$, 
${\cal R}_{s}$ being the SMS radius) by
 removal of stars in the central regions of a stellar system (Peebles, 1972).
 A drift of stars occurs in the centre of the stellar system in order to
 recover the equilibrium. In the special case of having a BH at the centre
 of the system, processes like tidal disruption lead to the destruction of
 the star. In this arena we have an outward energy flux created by the sinking
 of stars via relaxation processes (local, two-body, small-angle gravitational encounters).

When we consider the general case of an SMS, the basic picture is the same,
 but some aspects vary; the removal of stars and the inward transport are
 due to a different process. The inward flux of energy is produced not only
 for the local, two-body, small-angle gravitational encounters, but also for
 dissipative processes. The effective sinking for stars is now related to gas-drag
 energy dissipation and can involve or not their actual physical destruction.
 A star moving through the cloud will quickly dissipate energy because of gas
 drag and then it will sink into the centre of the SMS. The main difference
 between a BH and an SMS is basically that the former produces a low-angular
 momentum star depletion on a crossing time, whereas the dense gas cloud or
 SMS does the same but in a dissipation time.

Since we want to analyse the effects on the dense stellar system arising from
 the presence of a central gas cloud we have to distinguish between those stars
 whose orbits are limited to the region where the SMS is located ({\sl confined 
stars}) and those stars in orbits which surpass the radius of the
 cloud ({\sl unconfined stars}).

When we talk about confined stars, the first steps in the evolution are
 determined by the energy dissipation given by the drag force that the 
dense gas cloud exerts on the individual stars. Stars lose velocity in
 their motion inside the cloud, they are slowed down by the gas and therefore
 they cede heat to the cloud.

The slowing down of the stars makes them become a more compact subsystem
 which will sink down to the centre of the cloud. The system becomes
 self-gravitating and we have a cusp in the stellar distribution. However,
 star-star interactions can play a decisive role in this point, since they
 yield a depletion in the number of confined stars, or we can have direct 
collisions between them and thus disruption or coalescence. This could avoid
 that a singularity in the core collapse crops up. We cannot exclude the
 exchange of mass between individual stars and the gas as another possible
 way to prevent the singularity, since this can yield the star disruption
 via stellar wind, or the creation of heavier stars which become a
 supernova (Da Costa 1979).

The consequent evolution of the confined system will be in part determined
 by the rate at which surrounding stars outside the gas cloud refill this
 confined-stars gap. The importance of the core collapse will also be a decisive 
point for the evolution.

Unconfined stars move on orbits extending larger than the SMS radius.
 However, they can suffer its influence out to a radius within which
 the presence of the SMS is effective. The idea is exactly the same as
 for the BH, for the SMS is a generalisation of the former case

\subsection{Kinetic energy dissipation}

The drag force that the individual stars suffer when they cross the SMS
 is given by the next equation estimated by Bisnovat'i-Kogan and S'un'aev (1972):

\begin{equation}
F_{D}=C_{D} S \rho_{sms} v_{*}^2.
\end{equation}
\noindent
$C_{D}$ is a numerical parameter of order unity, $\rho_{sms}$ is the mean density
 of the gas cloud, $v_{*}$ is the velocity of the stars and $S$ is the cross section of
 the stars, $S=\pi r_{*}^2$.
 In case of a supersonic motion of the star the force $F_{D}$ can be
 interpreted as caused by the ram pressure (pressure difference) originating
 at a bow shock in front of the moving star. Due to the physical shock
 conditions one can show in such a case $C_{D} \approx 4 $ (Courant \& Friedrichs 1998). 
 
Suppose that the star crosses the SMS from one extreme to the opposite, i.e.
along its diameter; thus, if ${\cal R}_{s}$ is the radius of the SMS,
the stellar energy dissipated during each passage
through it is

\begin{equation}
\triangle E_{D}=F_{D} \cdot 2{\cal R}_{s}.
\end{equation}

The orbits of the stars within $r_{h}$ will be elliptic shaped with one
 focus at the SMS centre. The semi-major orbit axis $a$ will shrink because
 of the drag force, driving the orbit directly into the SMS. The average energy
 dissipation rate is
\begin{equation}
- \frac {dE}{dt} = \frac {\triangle E_{D}}{T} = \frac {2C_{D} \rho_{sms} \pi
 r_{*}^2 G{\cal M}_{s}}{ \pi \sqrt {4a^3/G{\cal M}_{s}}},
\end{equation}
\noindent
where $T$ is the period and ${\cal M}_{s}$ the mass of the SMS.

\subsection{Loss-cone stars velocity field distribution function}

 It stands to reason that at distances much larger than the SMS radius 
 ($r \gg {\cal R}_{s}$)
 the star field velocity distribution has a Maxwellian shape: 
\begin{equation}
f(v_{t},v_{r})= \frac{\rho(r)}{(2\pi)^{3/2}\sigma_{r}\sigma_{t}^2} \exp
 (-\frac {v_{r}^2}{2\sigma_{r} ^2})\exp (-\frac {v_{t}^2}{2\sigma_{t} ^2}).
\end{equation}
In order to get the density of stars within the loss-cone we have to compute
 the following integral:
\begin{equation}
\rho_{lc}(r)= \int_{-v_{r}|_{\rm max}}^{0} \int_{-v_{lc}}^{+v_{lc}} f(v_{r},
v_{\theta},v_{\phi}) \mathrm{d}v_{r}{d}v_{\theta}{d}v_{\phi}.
\end{equation}
\noindent
Taking into account that $\mathrm {d}v_{\theta}{d}v_{\phi}=2\pi v_{t}
 {d}v_{t}$ and that $f(r,v)=f(r, -v)$,
\begin{equation}
\rho_{lc}(r)=4 \pi \rho(r) \int_{0}^{v_{r}|_{\rm max}} \mathrm{d}v_{r}
 \int_{0}^{v_{lc}(r,v_{r})} f(r,v_{r},v_{t})v_{t} \mathrm{d}v_{t}.
\end{equation}
In this expression, the maximal radial velocity is $v_{r}|_{\rm max}=
v_{\rm escape}=\sqrt{2 \phi(r)}$ and the potential is the sum of both,
 the supermassive star and the stellar system potential,
\begin{equation}
\phi(r)=\phi_{sms}(r)+\phi_{*}(r).
\end{equation}

The integral happens to be analytical, and it yields the following result:
\begin{equation}
\rho_{lc}(r,v_{r})=\rho(r) \cdot \{ \alpha - \zeta \cdot \beta \cdot \psi \},
\end{equation}

\noindent
where 

\begin{displaymath}
\alpha \equiv {\rm erf} (\sqrt {\phi(r)/\sigma_{r}^2})
\end{displaymath}
\begin{displaymath}
\beta \equiv \exp \Bigl( - {{\cal R}_{s}^2 \over r^2-{\cal R}_{s}^2} {2\triangle
 \phi \over \sigma_{t}^2} \Bigr)
\end{displaymath}
\begin{displaymath}
\psi \equiv {\rm erf} ( \frac {1}{\zeta} \sqrt { \phi(r)/ \sigma _{r}^2} )
\end{displaymath}
\begin{displaymath}
\zeta \equiv  [\frac{(r^2-{\cal R}_{s}^2)\sigma_{t}^2}{(r^2-{\cal R}_{s}^2)\sigma_{t}^2+{\cal R}_{s}^
2\sigma_{r}^2}]^{1/2}
\end{displaymath}
\begin{displaymath}
\triangle \phi \equiv \phi({\cal R}_{s})-\phi(r).
\end{displaymath}

Since we are working with a Gaussian function whose width is $\sigma$,
 the contributions of velocities $v_{r}>2\sigma_{r}$ to the total mass
 are small and therefore negligible. In the practice it means that we can
 approximate the integral by $v_{r}=2\sigma_{r}$. In such a situation, the
 loss-cone adopts an easy geometrical form: an open cylinder in the $-v_{r}$
 direction with a radius $v_{lc}$.
\noindent
The resulting integral yields

$$\frac {\rho_{lc}(r)}{\rho(r)}= \Bigg[ 1- \exp \Bigg\{ - \frac{{\cal R}_{s}^2}
 {r^2-{\cal R}_{s}^2} \frac{2\sigma_{r}^2 + \triangle \phi (r)}{\sigma_{t}^2}\Bigg\}
 \Bigg]  
$$
\begin{equation}
\hskip2cm \times {\rm erf} \Bigg( \sqrt {\frac {\phi (r)}{\sigma_{r}^2}} \Bigg)
\end{equation}
\noindent
We use for the stellar system a Plummer model. Thus,
\begin{equation}
\phi(r)=\frac {G{\cal M}_{s}}{r} + \frac{GM_{*}}{\sqrt{r^2 + r_{c}^2}}.
\end{equation}

\noindent
Therewith, the resulting expression is

\begin{equation}
\frac {\rho_{lc}(r)}{\rho(r)}= (1-\exp {A})\cdot {\rm erf}(B),
\end{equation}

\vskip0.1cm
\noindent
where

\begin{displaymath}
A \equiv - \frac{{\cal R}_{s}^2} {r^2-{\cal R}_{s}^2} ~ \frac {1}{\sigma_{t}^2}(C)
\end{displaymath}
\begin{displaymath}
B \equiv  \frac{1}{\sigma_{r}} \sqrt {\frac{G{\cal M}_{s}}{r} + \frac{GM_{*}}
{\sqrt{r^2 + r_{c}^2}}}
\end{displaymath}
\begin{displaymath}
C \equiv  2\sigma_{r}^2 + \frac{G{\cal M}_{s}}{{\cal R}_{s}} - \frac{G{\cal M}_{s}}{r}+
\frac{GM_{*}}{\sqrt{{\cal R}_{s}^2+r_{c}^2}} - \frac{GM_{*}} {\sqrt{r^2+r_{c}^2}}.
\end{displaymath}
\vskip0.2cm
The velocity vectors of all stars which belong to a given (fixed) phase
 space density $f=f_{0}={\rm const}$ 
shape an ellipsoid whose two tangential and one radial major axes have lenghts
 equal to the velocity dispersion:
 $\sigma_{\theta}=\sigma_{\phi}$, and $\sigma_{r}$ (Frank \& Rees 1976). If one
 uses $f_{0}=f(\sigma_{r},\sigma_{\theta},
 \sigma_{\phi})$, the surface of the ellipsoid $A= \pi \sigma_{\theta} \sigma_{\phi}
 \sigma_{r}$ is a measure for the available velocity space.

A cone of angle $\theta_{lc}$,
\begin{equation}
\theta_{lc}:={\rm arcsin} ~(\frac{v_{lc}}{\sigma_{r}}),
\end{equation}
cuts out a segment of the foregoing velocity ellipsoid's $A$ surface of
 corresponding fraction surface $A_{lc}$, 
\begin{equation}
A_{lc} \approx \pi \theta_{lc}^2,~ (\theta_{lc}<<\pi).
\end{equation}
We can then define the ratio $\Omega':=A_{lc}/A \approx \theta_{lc}^2/4$,
 which is a measure for the loss-cone size. 

Spurzem (1988) proofs that, with the assumption that $v_{lc}$ does not
 depend on $v_{r}$ any more and taking into account that with a Schwarzschild-Boltzmann
 distribution we can reduct the loss-cone momenta to elementary Gaussian
 error functions, the quantity
\begin{equation}
\Omega:=\frac{\rho_{lc}}{\rho}
\end{equation}
yields in first order
\begin{equation}
\Omega \approx \frac {v_{lc}^2}{4 \sigma_{\theta}^2} \approx \frac 
{\theta_{lc}^2}{4}.
\end{equation}

This is the connection between the preceding simple picture of the loss-cone and 
the definition of $\Omega$ in the velocity space: For a Schwarzschild-Boltzmann 
distribution we can find out that, at first order, $\Omega$ is of the same size as
 $\Omega'$.

\subsection{Isotropy and anisotropy in the stellar system}

We introduce now an {\sl isotropy ratio} in order to study the different
 possible situations for the stellar distribution: The tangential velocity
 dispersion is $\sigma_{t}^2=\sigma_{\phi}^2+\sigma_{\theta}^2$; in case of 
isotropy, $\sigma_{r}^2=\sigma_{\phi}^2+\sigma_{\theta}^2$, then $\sigma_{t}^2=
2\sigma_{r}^2.$

Now we define the ratio $R:=2\sigma_{r}^2/ \sigma_{t}^2$. According to this 
definition, 
$R=1$ for the isotropic case. The corresponding values of $R$ for radial and 
tangential 
anisotropy can be obtained bearing in mind that $\sigma^2=\sigma_{r}^2+\sigma_{t}
^2=\sigma_{t}^2
 ( R/2 +1)$, $\sigma_{t}=\sigma / \sqrt{R/2+1}$ and $\sigma_{r}=\sigma_{t}\cdot
 \sqrt{R/2}.$

We have the loss-cone star density as a function of the supermassive star radius,
 ${\cal R}_{s}$.
 This does not provide much information, since in principle this radius could have 
any size;
 we do not have a criterion for it yet. Instead, what does make sense is to
 express this
 loss-cone star density in terms of the supermassive star stability, which
 is something has been studied in detail (Fuller \& Woosley, 1986). 

From Chandrasekhar (1964), instability sets in when the radius of the star
 ${\cal R}_{s}$ is less 
than a critical radius ${\cal R}_{s}^{\rm crit}$. He shows that if the ratio of 
specific heats 
$\gamma=C_{p}/C_{v}$ exceeds 4/3 only by a small amount, then dynamical
 instability 
will occur if the mass contract to the radius ${\cal R}_{s}^{\rm crit}$
\begin{equation}
{\cal R}_{s}^{\rm crit}=\frac {K}{\gamma - \frac{4}{3}} \Bigg(\frac{2G{\cal M}_{s}}{c^2}
 \Bigg).
\end{equation}

Thus, we introduce the {\sl stability coefficient} $\delta:={\cal R}_{s}/{\cal R}_{s}^
{\rm crit}$. 
We just have to substitute ${\cal R}_{s} = \delta \cdot {\cal R}_{s}^{\rm crit}$ in 
the loss-cone star
 density formula and vary $\delta$ instead of ${\cal R}_{s}$. 
\subsection{Connection at the influence radius}

Since we are interested in the diffusion angle, we now derive two expressions 
for it, within the influence radius of the SMS and outside it. 
For this aim we look at the dynamical and relaxation time at this radius. 

Within the influence radius the velocity dispersion is $\sigma(r)= 
\sqrt{G{\cal M}_{s}/{r}}$. This is just an approximation which we make
 here for simplicity, since our radii are $r<r_{h}$. To include
 $r\ll r_{h}$ we need a better model, which can
 only be obtained by  numerical solution of the equation of Poisson.
 Outside the influence radius, we use a modified Plummer
 model for the velocity dispersion, since we have to match both solutions, 
within and outside 
the SMS influence radius, since we have to look for a velocity dispersion 
connection; otherwise 
we get artificial, non-physical ``jumps'' in the plots for $r \approx r_{h}$.
 This can be performed by adding a factor $\alpha$ to the Plummer velocity
 dispersion expression, which we determine by requiring both velocities dispersions
 to be equal at the influence radius, 
$\sigma(r_{h})|_{r<r_{h}} \equiv \sigma(r_{h})|_{r>r_{h}}$. Thus, 
\begin{equation}
\sqrt{\frac{G{\cal M}_{s}}{r_{h}}}= \alpha \cdot \sqrt{ \frac{GM_{*}}{6r_{c}}} ~
 \bigg(1+\frac{r_{h}^2}{r_{c}^2} \bigg)^{-1/4}.
\end{equation}
\noindent

Therefore, the velocity dispersion outside the influence radius is:

\begin{equation}
\sigma(r) =\sqrt{ \frac{G{\cal M}_{s}}{r_{c}r_{h}}} \cdot \bigg[ \frac {r_{c}^2}
{r_{c}^2+r^2} (r_{c}^2+r_{h}^2) \bigg]^{1/4}.
\end{equation}

Note that $\alpha$ is necessary because our velocity dispersion is approximate
 for $r<r_{h}$, but not for $r \ll r_{h}.$

For the dynamical and relaxation times we can get their corresponding 
expressions thanks to expressions (6) and (7).
\section{Mass accretion rates}

The rate of stars plunging onto the central SMS is given by two different 
formulae, depending on whether or not there is a crossing point for the 
$\theta_{lc}$, $\theta_{D}$- plot against the radius, {\em a critical radius}.
 If we find that the curves happen to cross, the mass accretion rate 
$\dot{M}$ has the expression $\dot{M}=M_{*}(r_{h})/t_{\rm relax}(r_{h})$ 
because the loss-cone will be depleted in a relaxation time, and the mass
 to take into account is that which lays within the critical radius. 
On the other hand, if there is no crossing point, this means that the 
loss-cone in not empty and in this situation we have to employ a rather
 different expression, we have to resort to the loss-cone star density 
expression to get the mass being accreted into the SMS. In this case the
 timescale of interest is the dynamical time, $\dot{M}=\Omega(r) 
M_{*}(r)/t_{\rm dyn}(r)$.

It may be asserted, nevertheless, that this formula is not completely correct because it is based on
 the stationary model,
 which supposes an empty loss-cone in the first case and a full loss-cone in 
the second case.
 We have to generalise it by means of a ``diffusion'' model (Spurzem 2000).
 We introduce the concept
 of the {\em filling degree k} of the loss-cone as follows: Let us conjecture that $f$
 is the unperturbed velocity distribution. If the loss-cone is empty and 
angular momentum diffusion is neglected,
 then $f=0$ inside the loss-cone and $f$ remains unchanged elsewhere in velocity space.
 Actually,
 this distribution function will have a continuous transition from nearly
 unperturbed values
 at large angular momenta towards a partially depleted value inside the
 loss-cone. We approximate
 this by a distribution function $f$ having a sudden jump just at the value
 $L_{\rm min}=m_{*} v_{lc}$ from an unperturbed value $f_{0}$, $f=k \cdot f_{0},
~{\rm with}~0 \le k \le 1$.
 Since we work with the hypothesis that some kind of stationary state is to be
 established
 in the limit $t \to \infty$, the filling degree is
\begin{equation}
k_{\infty}=\frac{ \nu (1+\nu)}{1+\nu(1+\nu)}.
\end{equation}
\noindent
In this expression $\nu \equiv { \theta_{D}^2}/{\theta_{lc}^2}.$
Then we have to multiply the accretion rates by this filling degree $k_{\infty}$.

The stellar mass within and outside the influence radius is
\begin{equation}
M(r,r_{\rm min})|_{r<r_{h}}=\frac{16\pi}{5} \rho (r_{h})~r_{h}^3\bigg
( \frac{r}{r_{h}} \bigg)^{5/4},
\end{equation}
\begin{equation}
M(r)|_{r>r_{h}}=M(r_{h})|_{r<r_{h}} + 4\pi \int_{r_{h}}^{r} \rho(r'){r'}^2dr'
\end{equation}
$$=\frac{16\pi}{5} \rho (r_{h})~r_{h}^3 + 4\pi \frac{M_{*}}{r_{c}^3} 
\bigg[ \frac{r^3}{ \big( 1+ {r^2}/{r_{c}^2}\big)^{3/2}}  - \frac{r_{h}^3}
{\big( 1+ {r_{h}^2}/{r_{c}^2}\big)^{3/2}}\bigg].$$

To get the total heating rates, we just have to compute the value of
\begin{equation}
L_{\rm heat,~all} = \bigg( \frac {\dot{M}}{m_{*}} \bigg)\cdot E_{\rm heat,~1*},
\end{equation}
\noindent
where $E_{\rm heat,~1*}$ is the heating for one star (during one crossing).
\section{Plots and results}
In this section we analyse the interaction rate of stars with the SMS by varying
 the parameters introduced in the former sections, namely $\delta$, the core radius $r_{c}$,
 the total stellar mass and the supermassive star mass itself. We suppose that the stars which
 conform the stellar system are solar-type stars. For all the plots we extend the radii down
 to 1.001 times the SMS radius, because we would run into a snag if we extended it within
 the SMS radius, since the proportions $\sigma^2 \propto 1/r$ and $\rho \propto r^{-7/4}$
 would be wrong and the loss-cone star density and therefore the loss-cone angle 
has been obtained considering a Maxwell-Boltzmann distribution. It would also be an
 error of the problem conception
 itself, because we are studying the non-confined stars and the loss-cone, and its
 definition does not make any sense for radii less than the ${\cal R}_{s}$. This explains
 the first inequality in ${\cal R}_{s}<r_{h}<r_{c}$, which is an exigency that we must follow
 unfailingly because, otherwise, the Plummer model, which we use for the stellar system,
 would not be a suitable solution for it in the case that $r_{h}<r_{c}$ is not satisfied.
 However, what we demand here is not a requirement of the physics of the problem, but a
 condition for the method being employed to solve it. Situations in which $r_{h}<{\cal R}_{s}$
 or $r_{h}>r_{c}$ have to be solved numerically for the equation of Poisson and the velocity
 distributions.

In order to estimate the heating rates for a single star crossing the SMS we have to plot
 out the mass accretion rate of this central massive object in the case that we have no 
crossing point for the loss-cone and diffusion angle curves, whereas if we have a critical
 radius we will have to compute the total stellar mass and dynamical time at this value.

In Fig. 1 we plot the velocity dispersion against the radius for a $10^3\msol$ SMS.
 We observe a typical
 power law of $\propto r^{-1/2}$ within the influence radius -which we represent by a vertical 
dashed line for all cases-
 because we have a cusp on velocities in this interval of radii. We have a nearly constant
 velocity for later radii which lie in the section of values close to the core radius. Then
 the velocity dispersion decays. The length of this nearly constant velocity section
 as well as the slope for the decay depends on the galactic nuclei in the galaxy. 

Regarding Fig. 2, we show the loss-cone normalised density difference between the 
isotropic and radially anisotropic cases for a $10^7\msol$ SMS. In
 this plot we can observe a bigger number of stars being accreted into the SMS for the
 radially anisotropic case,since a radial orbit means a lower angular momentum 
and thus it is more probable
 that the star sinks into the central object, and vice-versa for a tangential orbit (Fig.3). 

As regards the loss-cone and diffusion angle plots, we examine two different cases in Fig. 3 and 4: 
a $10^4\msol$ and a $10^7\msol$ SMS. One may observe that for the first one we find a critical radius, 
whereas for the latter one the curves do not intersect. It is also interesting to find 
out which angle is bigger and where, since for a $\theta_{lc} \gg \theta_{D}$ 
we have an almost empty loss-cone, because it is quickly depleted and the stars 
replenish it very slowly; the opposite case, $\theta_{lc} \ll \theta_{D}$, implies
 that the loss-cone is full. 

We get a maximum $\dot{M}$ at about a core radius in the mass 
accretion rates plot for a $10^7\msol$ SMS, because the biggest contribution
 of stars being accreted into the SMS lies at this radius. Figure 5 shows
 an irregularity
 at the influence radius, because the loss-cone and diffusion angle plots show us that the
 former happens to be always bigger than this one and thus we have to apply the approximation
 commented in the foregoing section.
\subsection{Heating rates. An estimation}

When we look at the $\dot{M}$ for the $10^3\msol$ and $10^4\msol$ SMS's, we obtain that
for the $10^3\msol$:
\noindent
 ${\dot{M}}_{\rm iso}=1.75 \times 10^{-13} ~ \msol/{\rm yr}$, 
\noindent
 ${\dot{M}}_{\rm tan}=1.66 \times 10^{-13} ~ \msol/{\rm yr}$
\noindent
 ${\dot{M}}_{\rm rad}=1.97 
\times 10^{-13}~ \msol/{\rm yr},$. 
\noindent
where the subscript ``iso'' stands for isotropy, ``rad'' for radial anisotropy
 and ``tan'' for tangential anisotropy. 

The case of $10^4\msol$ is similar to the last one. For the $10^7\msol$
 SMS, we will select the $\dot{M}$ corresponding to the core radius, since the most
 important contribution is reached there: $\dot{M}|_{\rm core} =
 10^{-2}~ \msol/{\rm yr}.$ To get the heating rates of these non-confined stars,
 we just have to compute
\begin{equation}
\dot{E}= \bigg( \frac{\dot{M}}{\msol} \bigg) \pi r_{*}^2 \rho_{sms}
 v_{*}^2 \cdot 2{\cal R}_{s}/{t_{\rm cross}},
\end{equation}
where $v_{*}^2=G{\cal M}_{s}/{\cal R}_{s}$, $t_{\rm cross}= 2{\cal R}_{s}/v_{*}$,
 $\rho_{sms}={\cal M}_{s}/(\frac{4 \pi}{3} {\cal R}_{s}^3).$
\noindent
For the SMS we have supposed, as a first approximation, a constant density.

The corresponding ${\cal R}_{s}$ are:

\vskip0.3cm
\noindent
$10^3\msol$: ${\cal R}_{s}=2.5 \times 10^{-9}$ pc

\noindent
$10^4\msol$: ${\cal R}_{s}=8 \times 10^{-8}$ pc

\noindent
 $10^7\msol$: ${\cal R}_{s}=2.5 \times 10^{-2}$ pc.

\vskip0.2cm
\noindent
The luminosities are:

\vskip0.3cm
\noindent
$L_{10^3}/\lsol =6.2 \times 10^3$

\noindent
$L_{10^4}/\lsol =1.17 \times 10^4$

\noindent
$L_{10^7}/\lsol =5.6 \times 10^5,$

\noindent
where $L_{10^i}$ stands for the $10^i\msol$ SMS luminosity.
It is not a surprise that these luminosities are not sufficient to
support quasar luminosities, which was known before. We confirm however
the earlier result by Langbein et al. (1990) with our more detailed, but
stationary loss-cone model, that the luminosities are large enough to
prevent for some time the relativistic collapse of a SMS in a galactic
centre; that was called the quasi-pile stage by Hara (1978).

\begin{figure}
\vspace{6.0cm}
\includegraphics{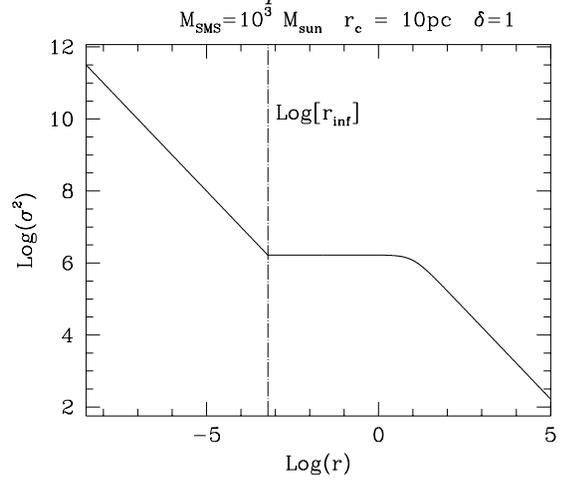}
\caption{The velocity units are $(m/s)^{2}$ and $r$ is expressed in pc. The influence radius is 
located at $6 \cdot 10^{-4}$pc. We show its logarithm in the plot with a vertical dashed
 line. The central velocity dispersion 
(the velocity dispersion at the influence radius) is $\sigma_{\rm central}=\sqrt{GM_{*}/(6r_{c})}=84$ km/s. ${\rm M}_{\rm SMS}$ stands for ${\cal M}_{s}$.}
 \end{figure}

\begin{figure}
\vspace{6.0cm}
\includegraphics{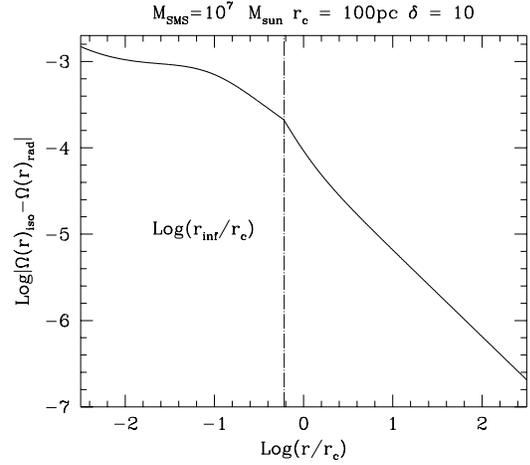}
\caption{$\Omega(r):=\rho_{lc}(r)/\rho(r)$ decays with the distance to the SMS. We have divided
 the radius by the core radius in order to normalise it. A bigger number of stars plunge onto the SMS
for the radially anisotropic case.}
 \end{figure}

\begin{figure}
\vspace{6.0cm}
\includegraphics{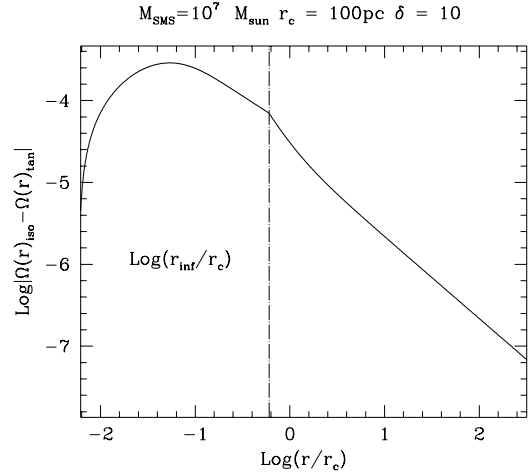}
\caption{The same as for Fig.2 but for the tangentially anisotropic case and a lower number of stars.}
 \end{figure}

\begin{figure}
\vspace{6.0cm}
\includegraphics{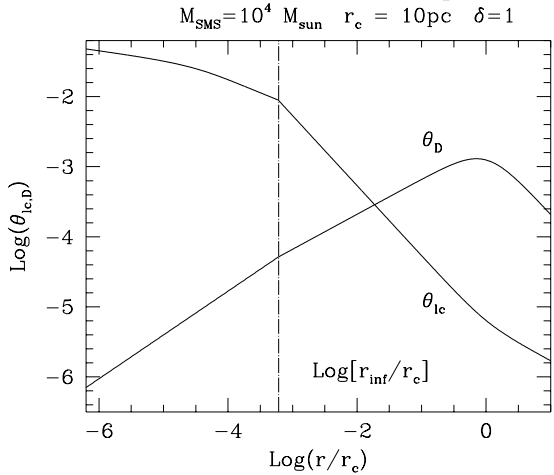}
\caption{$\theta_{lc}=2 \cdot \sqrt{\Omega(r)}$, $\theta_{D}= \sqrt{ t_{\rm dyn} / t_{\rm relax}}$.
 For this mass the two curves cross at the critical radius. For radii smaller than that, the loss-cone
 angle is bigger than the diffusion angle, and this implies that the loss-cone is empty. From the
 $r_{\rm crit}$ onwards it is no longer empty, for $\theta_{D}>\theta_{lc}$. With the crossing
 point we can work out the accretion rate and find out the differences depending on whether we
 consider an isotropic situation for the stellar velocity distribution function or an anisotropic
 one, distinguishing between a radial or tangential anisotropy. If we set this case against the $10^3\msol$ we do not find big differences.}
 \end{figure}

\begin{figure}
\vspace{6.0cm}
\includegraphics{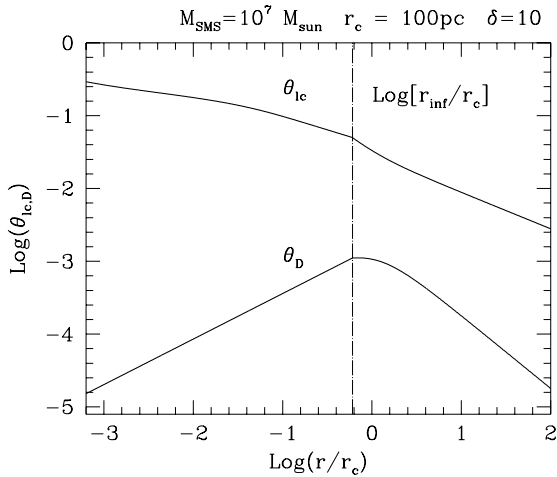}
\caption{For higher masses, such as $10^7\msol$, the loss-cone is empty.
 We get no critical radius.}
 \end{figure}

\begin{figure}
\vspace{6.0cm}
\includegraphics{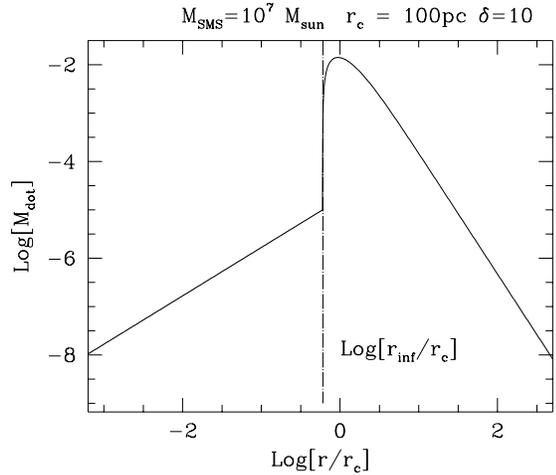}
\caption{The ``jump'' is due to the approximation we made in order to get
 the accretion rates. The maximum is reached at a distance which corresponds to the core radius.}
 \end{figure}

\section{Conclusion and Discussion}

We have revisited the classical loss-cone semi-analytic theory invented
by Frank \& Rees (1976) for star accretion onto central supermassive black
holes in galactic nuclei and star clusters and extended by da Costa (1981)
for the case of stars on radial orbits being trapped by star-gas interactions
in a central supermassive star-gas system. In Langbein et al. (1990) such
model was included in time-dependent, spherically symmetric models of
star-gas systems in galactic nuclei. Though highly idealized we think such
configurations are still worth a study to understand the physical processes
at work in the early formation phase of massive galaxies with formation of
central black holes. Notions such as the critical radius in the classical
work, where the loss-cone star accretion becomes important and flattens out the
cusp density profile turn out naturally in our model without any ad hoc
assumptions. While in this research note we keep the stellar background system
fixed and develop the old ideas in an up-to-date form, we will present in
subsequent papers their inclusion into a self-consistent dynamical
model of relaxing star clusters with central black hole, first
in spherical symmetry (using high resolution anisotropic gas models as described
in Louis \& Spurzem 1991, Spurzem 1994 and Giersz \& Spurzem 1994), later
going to more general symmetries or triaxial systems. Recent dramatic
improvements of the observational situation and the demography of black
holes in nuclei demand such progress in dynamical modelling, which is yet
surprisingly poor (see e.g. Richstone 1998). It is clear that in a proper
cosmological context many complications occur for simplified modelling: the
whole system is embedded in a collisionless dark halo with disputed central
density profile, it is non-stationary due to a sequence of merger events
in hierarchical structure formation, with the possible formation of binary or
multiple black holes and perturbations of various kinds will even cause
a single black hole not to be fixed in the centre. Despite of all that
have begun our work at the point where we think present (astro)physical
understanding and modelling comes to its limits, and this is the case for
a spherical dense large $N$ star cluster, suffering from relaxation
and star accretion around a fixed massive black hole.

 \section*{Acknowledgements}

P. Amaro-Seoane would like to thank Francine Leeuwin for her inestimable help and useful
 discussions. This work has been supported by Sonderforschungsbereich (SFB) 439
 ``Galaxies in the Young Universe'' of German Science Foundation (DFG) at the University
 of Heidelberg, performed under the frame of the subproject A5.

 \end{document}